\magnification=1200                                                             
                             
\font\titre=cmcsc10 at 16pt                                                     
                                                               
\font\small=cmr10 
\parskip .3cm                                                                   
\baselineskip=.50cm                                                             

\vsize=200truemm                                                                               
\voffset=1truecm                                                               
\line{\hfill ISN -96.99}
\vfill                                                        
\vskip.9cm                                                                     
\centerline{\titre  Tetraquarks with colour-blind
forces}                        
\vskip .25cm                                                                     
\vfill                                                        
\centerline{\titre in chiral quark models}                            
\vskip .7 cm                                                                    
\centerline{{\bf S. Pepin}\footnote{$^1$}{\small e-mail: {\tt u2162et@vm1.ulg.ac.be}}   
 and {\bf Fl. Stancu}\footnote{$^2$}{\small e-mail: {\tt u216207@vm1.ulg.ac.be}}} 
\centerline{\small Universit\'e de Li\`ege}                                       
\centerline{\small Institut de Physique, B.5, Sart Tilman}                      
\centerline{\small B--4000 Li\`ege 1, Belgium}                                  
\vskip .2cm                                                                     
\centerline{{\bf M. Genovese}\footnote{$^3$}{\small e-mail: {\tt
genovese@isnmv1.in2p3.fr}}
\footnote{$^4$}{\small Supported by the EU Program ERBFMBICT 950427}
 and 
{\bf J.-M. Richard}\footnote{$^5$}{\small e-mail:{\tt jmrichar@isnnx2.in2p3.fr}}
}                 
\centerline{\small Institut des Sciences Nucl\'eaires}                          
\centerline{\small Universit\'e Joseph Fourier--IN2P3-CNRS}                     
\centerline{\small 53, avenue des Martyrs, F-38026 Grenoble Cedex,              
France}                                                                         
                                                                                
\vfill                                                        
\vskip .5cm                                                                     
\centerline{\bf Abstract}                                                       
{\baselineskip=.45cm \parindent=2cm\narrower\noindent\small 
We  discuss the stability of  multiquark systems within the recent model
of Glozman {\sl et al.\/} where the chromomagnetic hyperfine interaction
is replaced by  pseudoscalar-meson exchange contributions. 
We find that such an
interaction binds a heavy tetraquark systems $QQ\bar q\bar q$
($Q=c,\,b$ and $q=u,\,d)$ by
$0.2-0.4\;$GeV. This is at variance with results of previous models where 
$cc\bar q\bar q$ is
unstable.\par}                                              
\par\noindent                                                                   
\vskip.4cm                                                                     
\vfill \eject                                                                                
\noindent The  existence of tetraquark hadrons ---two quarks and two
antiquarks--- has been raised about twenty years ago by Jaffe [1]
and has been studied within a variety of models. The MIT bag study
indicated the presence of a dense spectrum of tetraquark states in
the light sector [1]. Later on, tetraquark systems have been
examined in potential models [2-4] and flux tube
models [5]. In particular the question of stability has been
raised, that is whether the tetraquark ground state lies below or above
the lowest $(q\bar q)+(q\bar q)$ threshold. Weinstein and Isgur
showed [2] that there are only a few weakly--bound states of
resonant meson--meson structure in the light
$(u,\,d,\,s)$ sector. On the other hand, no bound state was found by
Carlson and Pandharipande [5] in their flux-tube model with
quarks of equal masses.\par
One important result in the MIT bag or potential models is that the
chromomagnetic interaction plays a crucial role [2,6] in
lowering the ground state energy of  a light system. Otherwise, in a
system of two heavy quarks and two light antiquarks
$QQ\bar q\bar q$ ($Q=c$ or $b$, $q=u,\,d$ or $s$) stability can be
achieved without spin--spin interaction, provided the mass ratio
$m(Q)/m(q)$ is larger [3] than about 15, which means that $Q$
must be at least a $b$-quark.\par
In the light sector, there are several candidates for non-$q\bar q$
states, but the experimental situation is not yet conclusive. For a
review, see for example Ref.~[7] and the last issue of Review of
Particle Properties [8]. In the heavy sector, experiments are
being planned at Fermilab and CERN, to search for new  hadrons  and
in particular for doubly charmed tetraquarks [9-11].\par
Recently, the baryon spectrum has been analysed by Glozman {\sl et
al.\/} [12,13] within a chiral potential model which includes
meson-exchange forces between quarks and entirely neglects the
chromomagnetic interaction. In view of its intriguing success in the
description of the baryon spectrum, it seems to us natural to apply
this model to multiquark hadrons with more than three quarks, and in
particular to tetraquarks.\par
A general Hamiltonian containing both chromomagnetic interaction and
me\-son-ex\-chan\-ge contributions has the form
$$                                                                              
\eqalign{                                                                       
 H=\sum_i{\vec{\rm p}_i^{\,2}\over 2m_i}&-{3\over16}                            
    \sum_{i<j} \tilde{\lambda}_i^{\rm                                           
c}\!\cdot\!\tilde{\lambda}_j^{\rm c}\,                                          
               V_{\rm conf}(r_{ij})\hfill\cr                                    
&{}-\sum_{i<j}\tilde{\lambda}_i^{\rm                                            
c}\!\cdot\!\tilde{\lambda}_j^{\rm c}\,                                          
\vec{\sigma}_i\!\cdot\!\vec{\sigma}_j\,V_{\rm g}(r_{ij})-                       
\sum_{i<j}\tilde{\lambda}_i^{\rm F}\!\cdot\! \tilde{\lambda}_j^{\rm             
F}\,                                                                            
\vec{\sigma}_i\!\cdot\!\vec{\sigma}_j \,V_{\rm F}(r_{ij}),                      
}                                                                               
\eqno(1)                                                                        
$$
where $m_i$ is the constituent mass of the quark located at
$\vec{\rm r}_i$; $r_{ij}=\vert \vec{\rm r}_j-\vec{\rm r}_i\vert$
denotes the interquark distance; $\vec{\sigma_i}$,
$\tilde{\lambda}_i^{\rm c}$,
$\tilde{\lambda}_i^{\rm F}$ are the spin, colour and
flavour  operators, respectively. Spin-orbit and
tensor components may supplement the above spin-spin forces for
studying orbital excitations. The potential in $H$ has three parts
containing the confining, the chromomagnetic and the meson-exchange
contributions, respectively.\par
The confining term $V_{\rm conf}$ usually consists of a Coulomb plus
a linear term,                               
$$                                                                              
V_{\rm conf}=-{a\over r}+b r+c.\eqno(2)                                           
$$
In the following, we shall use either the very weak linear potential
of Glozman {\sl et al.\/} [13] corresponding to
$$
(C_1)\qquad a=c=0, \qquad\hbox{and}\qquad b=0.01839\;{\rm GeV}^2,
\eqno(3)
$$
or  a more conventional  choice
$$
(C_2)\qquad a=0.5203, \qquad b=0.1857\;{\rm GeV}^2,
\qquad c=-0.9135\;{\rm GeV},\eqno(4)
$$
which has already been used in the study of tetraquarks by
Silvestre-Brac and Semay [4].\par
The third term in $H$ is often understood as the chromomagnetic
analogue of the Breit--Fermi term of atomic physics. For mesons,
$\tilde{\lambda}_1^{\rm c}\!\cdot\!\tilde{\lambda}_2^{\rm c}=-16/3$,
and a positive $V_{\rm g}$, as in the one-gluon-exchange model, shifts
each vector meson above its  pseudoscalar partner, for instance
$D^*>D$ in the charm sector. For baryons, where
$\tilde{\lambda}_1^{\rm c}
\!\cdot\! \tilde{\lambda}_2^{\rm c}=-8/3$ for each quark pair,  such
a positive $V_{\rm  g}$ pushes the spin 3/2 ground states up, and the
spin 1/2 down, for instance $\Delta >N$. In Ref.\ [4], the following
radial shape has been used
$$
V_{\rm g}={a\over\mathstrut m_im_jd^2}
{\exp-r/d\over\mathstrut r},\eqno(5) 
$$
with the same value of $a$ as in Eq.~(4) and $d=0.454\;$GeV$^{-1}$.
 \par
The last term of $H$ corresponds to meson exchange, and an explicit
sum over F is understood. If the system contains light quarks only (as in 
Refs. [12,13]),
the sum over F runs from 0 to 8 ($1-3\rightarrow\pi$, $4-7\rightarrow
K$, $8\rightarrow\eta$  and  $0\rightarrow\eta'$). If a heavy flavour
is incorporated,  a phenomenological extension from 
SU(3)$_{\rm F}$ to
SU(4)$_{\rm F}$ would further extend the sum to  ${\rm F} = 9-12$
corresponding to a $D$-exchange, ${\rm F} = 13-14$ to a
$D_s$-exchange and ${\rm F} = 15$ to an $\eta_c$-exchange. If
$V_{\rm F}=0$, one recovers a standard constituent quark
model.  The radial form of $V_{\rm F}\neq0$ is
derived from the usual  pion-exchange potential which contains a
long-range part and a short-range one
$$                                                                              
\sum_{i<j}\vec{\tau}_i\!\cdot\!\vec{\tau}_j\,\vec{\sigma}_i\!\cdot\!
\vec{\sigma}_j                                                                    
{g^2\over 4\pi}{1\over 4m^2}\left[\mu^2{\exp(-\mu r_{ij})\over                  
r_{ij}}-4\pi\delta^{(3)}(r_{ij})\right],\eqno(6)                                
$$
where $\mu$ is the pion mass. A coupling constant $g^2/4\pi=0.67$
at  the quark level corresponds to the usual strength
$g_{\pi N\!N}/4\pi\simeq14$ for the Yukawa
tail of the nucleon--nucleon ($N\!N$)
potential.
\par                                                            
When constructing $N\!N$ forces from meson exchanges, one disregards 
the short-range term in Eq.~(6), for it is hidden by the hard core, 
and anyhow the potential in that region is parameterized empirically. 
Similarly, when  T{\"o}rnqvist [14], Manohar and
Wise [15] or Ericson and Karl [16]
considered pion exchange in multiquark states, they had in  mind the
Yukawa term $\exp(-\mu r)/r$ acting between two well-separated quark
clusters. For similar  reasons, Weber {\sl et al.\/}[17],  in
their model with hyperfine plus pion-exchange interaction ignored  the
delta-term too. Therefore it is  somewhat of a surprise to see the
delta-term of Eq.~(6) taken seriously, and with an {\sl ad-hoc\/}
regularisation playing a crucial role in the quark
dynamics [12,13]. This regularised form
is [13]
$$
V_\mu=\Theta(r-r_0)\mu^2{\exp(-\mu r)\over
r}-                                 
{4\epsilon^3\over\sqrt\pi}\exp(-\epsilon^2(r-r_0)^2),                                    
\eqno(7)
$$
with the Yukawa-type part cut off for $r\le r_0$, where 
$r_0=2.18\;$GeV$^{-1}$, $\epsilon=0.573\;$GeV, and
$\mu= 0.139$ GeV for $\pi$, $0.547$ GeV for $\eta$ and $0.958$ GeV for 
$\eta '$. \par
Incorporating in $H$ both light quarks and charm (one can add bottom  similarly) and
working out the flavour matrix elements  of the mesons-exchange 
interaction between two quarks (or antiquarks), one obtains:
\catcode`\@=11
\def\system#1{\left\{\null\,\vcenter{\openup1\jot\m@th
\ialign{\strut\hfil$##$&$##$\hfil&&\enspace$##$\enspace&
\hfil$##$&$##$\hfill\crcr#1\crcr}}\right.}
\catcode`\@=12
$$
\left\langle
V_{i,j}\right\rangle=
\vec{\sigma}_i\!\cdot\!\vec{\sigma}_j\,
\openup 1mm\system{
&V_\pi&+&\displaystyle{1\over3}&V_\eta^{uu}&+&\displaystyle{1\over6}&V_{\eta_c}^{uu}
&&&     &;\quad [2]_{\rm F}&\  I&=&1\cr
2&V_K&-&\displaystyle{2\over3}&V_\eta^{us}&+&2&V_{D}^{uc}&+&&V_{D_s}^{sc}  
      &;\quad [2]_{\rm F}&\   I&=&1/2\cr
&&&\displaystyle{4\over3}&V_\eta^{ss}&+&\displaystyle{3\over2}&V_{\eta_c}^{cc}
&&&       &;\quad [2]_{\rm F}&\   I&=&0\cr
{}-2&V_K&-&\displaystyle{2\over3}&V_\eta^{us}&-&2&V_{D}^{uc}&-&&V_{D_s}^{sc}
         &;\quad [11]_{\rm F}&\   I&=&1/2\cr
{}-3&V_\pi&+&\displaystyle{1\over3}&V_\eta^{uu}&+&\displaystyle{1\over6}&V_{\eta_c}^{uu}
&&&           &;\quad [11]_{\rm F}&\   I&=&0\cr
}         \eqno(8)
$$
for ${\rm F}=1,\ldots15$, and 
$$\langle
V_{ij}\rangle={2 \over 3} \vec{\sigma}_i\!\cdot\!\vec{\sigma}_jV_{\eta'}
\eqno(9)$$ 
for ${\rm F}=0$.
This is an extension of Eq.~(3.3) of Ref. [12] from SU(3) to SU(4). 
Here $I$ is the isospin and in each case it is specified 
whether the pair is in a symmetric $[2]_{\rm F}$ state or in an 
antisymmetric $[11]_{\rm F}$ state. Actually little 
$u\bar u$ or $d\bar d$ mixing is expected in $\eta_c$ so that the
contribution of $V_{\eta_c}^{uu}$ can be safely
neglected. Moreover when the meson mass $\mu$ reaches
values of a few GeV as for $D$ or $\eta_c$ the two terms in Eq.~(6)
basically cancel each other so the expressions (7)  reduce
practically to their SU(3) form [12].  This is in agreement
with Ref.~[18] where it has been explicitly shown that the dominant
contribution to the $\Sigma_c$ and $\Sigma_c^*$ masses is due to
meson exchange between light quarks and the contribution of the
matrix elements with the $D$ ($D_s$) and $D^*$ ($D^*_s$) 
quantum numbers (which are evaluated phenomenologically, 
fitting the mass difference $\Sigma_c - \Lambda_c$) 
play a minor role. In the following
numerical calculations we will neglect the exchange of heavy mesons.\par
The remaining parameters are the quark masses. They are indicated in
Table 1, in conjunction with the two choices. The light quark masses
$m=m_u=m_d$ are from the corresponding literature [13,4].
The heavy quark mass $m_Q=m_c$ or $m_b$ is adjusted to reproduce the
experimental average mass $\overline{M}=(M+3M^*)/4$ of $M=D$ or
$B$ mesons. The meson mass is obtained from a trial wave function of
type $\phi\propto\exp(-\alpha r^2/2)$, with $\alpha$ as a variational
parameter. It has been checked that the error never exceeds a few
MeV with respect to the exact value. The variational approximation
is retained for consistency with the treatment of 3- and 4-body
systems discussed below.\par
\vskip .5cm {
\baselineskip=.15cm \parindent=2cm\narrower\noindent
 {\bf Table 1.} Quark masses and average heavy meson masses
$\overline{M}=(M+3M^*)/4$ ($M=D,\,B$) used for the potentials
$(C_1)$ and $(C_2)$. The units are GeV.\par
}
\vskip -.3cm
$$\vbox{\def\cc#1{\hfill\kern .7em #1\kern .7 em \hfill}
\def\tvi{\vrule height 12pt depth 5pt width 0pt}
\offinterlineskip
\halign{
\cc{#}&\cc{#}&\cc{#}&\cc{#}&\cc{#}&\cc{#}\cr
\noalign{\hrule}
\tvi Model& $m$ & $m_c$ & $m_b$ & $\overline{D}$ & $\overline{B}$ \cr
\noalign{\hrule}
\tvi $(C_1)$ & $0.340^{a)}$ & $1.350\phantom{^{b)}}$ & $4.660\phantom{^{b)}}$ & 
2.001 & 5.302 \cr
\tvi $(C_2)$ & $0.337^{b)}$ & $1.870^{b)}$ & $5.259^{b)}$ &  2.006 & 5.350 \cr
\noalign{\hrule}
}
\vfill}
$$
\vskip -.2cm
\line{\hglue 2cm ${}^{a)}$ Ref.~[13]\hskip1cm${}^{b)}$ Ref.~[4]\hfill}
\par
\vskip 1cm
We now briefly discuss the baryons. In the model of Glozman {\sl et
al.\/} the explicit form of the Hamiltonian integrated in the
spin--flavour space is :
$$                                                                              
H=H_0+{g^2\over48\pi m^2}                                                       
\left\{                                                                         
\normalbaselineskip=20pt                                                        
\matrix{                                                                        
&\!\!\!\!\!15V_\pi-V_\eta-2\left({g_0/                                          
g}\right)^2V_{\eta'}\quad\hbox{for}\quad
N\cr                                                                                                                   
&\!\!\!\!\!\phantom{1}3V_\pi+V_\eta+2\left({g_0/g}\right)^2V_{\eta'}
\quad\hbox{for}\quad\Delta\cr                                                     
}\right.                                                                                                                                                
\eqno(10)$$                                                                              
with                                                                            
$$                                                                     
H_0= 3m+\sum_i{\vec{\rm
p}_i^{\,2}\over                                 
2m}+{b\over2}\sum_{i<j}r_{ij},                                                                                            
\eqno(11)                                                                        
$$ 
where $g^2/4\pi = 0.67$, $(g_0/g)^2 = 1.8$ and $V_{\eta}
= V_{\eta}^{uu}$ of Eq.~(8).
\par
We have performed variational estimates with a wave
function $\phi \propto \exp(-\alpha (\rho^2 + \lambda^2)/2)$,
where $\vec{\rho}= \vec{\rm r}_2  - \vec{\rm r}_3$ ,
$\vec{\lambda}=(2\vec{\rm r}_1-\vec{\rm r}_2-\vec{\rm r}_3)/\sqrt{3}$,
and reproduced the results of the more elaborate Faddeev
calculations of Ref.~[13].  When the meson--exchange terms are switched
off, the $N$ and $\Delta$ ground states are degenerate at
1.63 GeV. When the coupling  is introduced, the wave
function is modified. For the nucleon, the                 
spin-independent part $H_0$ of the Hamiltonian 
gives a contribution of $2.11\;$GeV, and
receives a large $-1.14\;$GeV correction from meson exchange. For
the $\Delta$ ground state, the contribution of $H_0$ and meson
exchange parts are $1.91\;$GeV and $-0.63\;$GeV,
respectively. Thus one ends up with a reasonable value for the
$\Delta-N$ splitting, close to $0.3\;$GeV.
\par
We have also calculated the ground state baryons of content
$cqq$ with a trial wave function $\phi \propto 
\exp(-(\alpha \rho^2+\beta\lambda^2)/2)$ and found 
$\Lambda_{c} = 2.32\;$GeV and $\Sigma_c = \Sigma^{*}_{c} =
2.48\;$GeV, close to the experimental values and consistent with the
findings of Ref.~[18], although the Hamiltonian, its treatment,
and the input parameters are somewhat different there.\par
Due to arguments at the beginning of this letter, here we discuss
tetraquarks containing heavy flavours, i.e. $QQ \bar q \bar q$, 
studying the most favourable
configuration $\bar 3 3,\,S=1,\,I=0$. This means $QQ$ is
in a  $\bar 3$ colour state and $\bar q\bar q$ in a 3 colour state.
The mixing with $6\bar6$ is neglected because one expects 
this plays a negligible role in deeply bound heavy
systems [3]. Then the Pauli principle requires $S_{12}=1$ for
$QQ$, and $S_{34}=0,\, I_{34}=0$ for $\bar q\bar q$, if the relative
angular momenta are zero for both subsystems. This gives a state of
total spin $S=1$ and isospin $I=0$. \par
The tetraquark Hamiltonian integrated in the colour--spin--flavour
space, and incorporating the approximations discussed in relation to
Eq.~(8), reduces to
$$H=2(m+m_Q)+{\vec{\rm p}_x^2\over m_Q}+{\vec{\rm p}_y^2\over m}
+{m+m_Q\over2mm_Q}\vec{\rm p}_z^2+\sum_{i<j}V_{ij},\eqno(12)
$$
where
$$
\eqalign{
&V_{12}={1\over2}\left(-{a\over r_{12}}+b\, r_{12}+c\right),\cr
&V_{ij}={1\over4}\left(-{a\over r_{ij}}+b\, r_{ij}+c\right),
\qquad i=1\;\hbox{or}\;2,\;j=3\;\hbox{or}\;4,\cr
&V_{34}={1\over2}\left(-{a\over r_{34}}+b\, r_{34}+c\right)
+9V_\pi-V_\eta-2V_{\eta'}.}\eqno(13)
$$
The momenta $\vec{\rm p}_x$, etc., are conjugate to the relative
distances $\vec{\rm x}=\vec{\rm r}_1-\vec{\rm r}_2$, 
$\vec{\rm y}=\vec{\rm r}_3-\vec{\rm r}_4$, and 
$\vec{\rm z}=(\vec{\rm r}_1+\vec{\rm r}_2-\vec{\rm r}_3-\vec{\rm
r}_4)/\sqrt2$. The wave function is parameterized as
$$
\psi\propto\exp[-(\alpha x^2 +\beta y^2+\gamma z^2)/2],
\eqno(14)
$$
and the minimization with respect to $\alpha,\,\beta$ and $\gamma$
leads to the results displayed in Table 2 for $Q=c$ and $b$. Columns
2 and 3 corresponds to results derived from Eqs.~(12) and (13). This
shows that both the $cc\bar q\bar q$ and $bb\bar q\bar q$ systems
are bound whatever is the potential, $(C_1)$ or $(C_2)$, provided
meson exchange is incorporated. This is in contradistinction to
previous studies based on conventional models where the flavour-independent 
confining potential is supplemented by one gluon exchange. This remark is
illustrated by column 4 which shows that $cc\bar q\bar q$ is unbound
in such a case.
\par
\vfill\eject
\vskip .5cm {\baselineskip=.15cm \parindent=2cm\narrower\noindent
 {\bf Table 2.}
Energy (GeV) $\Delta E=QQ\bar q\bar q-2(Q\bar q)$ for 
heavy te\-tra\-quarks with $Q=c$ or $b$, $q=u,\, d$ in three 
different models with parameters defined in the text 
(OME = one meson exchange).\par}
$$\vbox{\def\cc#1{\hfill\kern .7em #1\kern .7 em \hfill}
\def\tvi{\vrule height 12pt depth 5pt width 0pt}
\offinterlineskip
\halign{
\cc{#}&\cc{#}&\cc{#}&\cc{#}\cr
\noalign{\hrule}
\tvi System& $(C_1)+{}$ OME & $(C_2)+{}$ OME & Ref.~[4]\cr
\noalign{\hrule}
\tvi $cc\bar q\bar q$ & $-0.185$ & $-0.332$ & $\phantom{-}0.019$ \cr
\tvi $bb\bar q\bar q$ & $-0.226$ & $-0.497$ & $-0.135$ \cr
\noalign{\hrule}
}
}
$$  

Considering this result one could rise the question if in the model
of Glozman  {\sl et al.\/} a
proliferation of multiquarks systems appears. We have therefore tried
to investigate $QQqqqq$ and $q^6$ systems as well. We have
proceeded analogously to the previous case using a Gaussian wave function.
In the case of $QQqqqq$, group--theory analysis [19] shows that the most
favourable configuration is the one where the light quark subsystem
has $S=1$, $I=0$,  corresponding to a global spin-flavour-averaged interaction 
$\langle V \rangle=10 V_{\pi} - 2/3 V_{\eta} - 4/3(g_0/g)^2
V_{\eta '}$. Our numerical calculation shows that this potential is largely
insufficient to bind the system. Also the $q^6$ system, the most favourable
configuration of which is $S=1$, $I=0$, leading to $ \langle V \rangle=
11 V_{\pi} - 5/3 V_{\eta} - 10/3(g_0/g)^2
V_{\eta '}$, is not sufficiently bound for being under 
the two baryons threshold.
                                          
It is also interesting to notice that the binding energy of $ c c \bar q \bar
q$ and $b b \bar q \bar q$ are nearly twice larger for the potential $(C_2$)
as compared to those of ($C_1$) and also much more different from each other.
The reason is that ($C_2$) contains a Coulomb part which binds more, heavier is
the system, leading thus to a larger separation among levels as well.
The potential ($C_2$) has been fitted to reproduce the $J/\Psi$ and the
$\Upsilon$ meson masses. It also gives overall good results both for other
mesons and baryons. By construction [12,13], the potential ($C_1$) was designed
and fitted to light baryons only. It is desirable to construct a chiral
potential model with a wider range of validity, covering the light and heavy
sector as well, and to apply it to the study of mesons and multiquarks systems.
Our results indicate that the chiral model of Glozman {\sl et al.\/} leads to
qualitatively different results for tetraquarks, with respect to commonly used
quark models. We hope that future experimental investigations might distinguish
among various approaches to quark dynamics. \par

\vfill \eject
\noindent {\bf Acknowledgements.} Useful discussions with L.Ya.~Glozman are gratefully 
acknowledged.

\vskip 1cm

\parindent=0cm\par                                                              
\vskip .5cm\noindent                                                            
{\bf References}\hfil\break                                                     
\parindent=1cm                                                                  
\frenchspacing                                                                  
\def\ref#1#2#3{\item{[#1]}\ {#2},\ {#3}}
\def\refs#1#2{ {#1},\ {#2}} 

\def\NPB{{Nucl. Phys.} B}
\def\NPA{{Nucl. Phys.} A}
\def\PLB{{Phys. Lett.}  B}
\def\PRL{{Phys. Rev. Lett.}}
\def\PRD{{Phys. Rev.} D}

\def\ZPC{{Z. Phys.} C}
\def\ZPA{{Z. Phys.} A}
\ref{1}{R.L. Jaffe}
{\PRD\ {\bf 15}, 267 (1977); {\bf 17}, 1444 (1978)}.
\ref{2}{J. Weinstein and N. Isgur}
{\PRD\ {\bf 27}, 588 (1983); {\bf 41}, 2236 (1990)}.
\ref{3}{S.Zouzou, B. Silvestre-Brac, C. Gignoux and J.-M.                     
Richard}{Z.~Phys. C {\bf 30}, 457 (1986)}.
\ref{4}{B. Silvestre Brac and C. Semay} {\ZPC {\bf 59}, 457 (1993);
{\bf 61}, 271 (1994)}.
\ref{5}{J. Carlson and V.R. Pandharipande}
{\PRD\ {\bf 43}, 1652 (1991)}.
\ref{6}{D.M. Brink and Fl. Stancu}{\PRD\ {\bf 49}, 4665
(1994).}              
\ref{7}{G. Karl}
{Int. J. of Mod. Phys. {\bf E1}, 491 (1992);
\NPA {\bf 558}, 113c (1993);}
\refs{N.A. T\"ornqvist}{Proc. of ``Int. Europh. Conf. on High Energy Phys.'',
Brussels (Belgium), edit. J. Lemonne et al., 84 (1995);} 
\refs{L. G. Landsberg}{Phys. of Atom. Nucl. {\bf 57}, 42  (1994).}
\ref{8}{Particle Data Group}
{\PRD\ {\bf 54}, 1 (1996).}
\ref{9}{M.A. Moinester}
{Tel Aviv University preprint TAUP 2255-95 (hep-ph/9506405)}, \ZPA (in press).
\ref{10}{D.M. Kaplan}
{Proc. Int. Workshop ``Production and Decay of Hyperons, 
Charm and Beauty Hadrons'',
Strasbourg (France) September 5--8, 1995}.
\ref{11}{COMPASS Collaboration (G. Baum et al.)}
{CERN--SPSLC--96-14, March 1996}.
\ref{12}{L.Ya. Glozman and D.O. Riska}
{Phys. Rep. {\bf 268}, 263 (1996).} 
\ref{13}{L.Ya. Glozman, Z. Papp and W. Plessas}
{\PLB\ {\bf 381}, 311 (1996)}.                                          
\ref{14}{N. T{\"o}rnqvist}{\PRL\ {\bf 67}, 556 (1991);\ZPC\ {\bf 61},
525 (1994).}
\ref{15}{A.V. Manohar and M.B. Wise}{\NPB\ {\bf 399}, 17 (1993).}
\ref{16}{T.E.O. Ericson and G. Karl}{\PLB\ {\bf 309}, 426
(1993).}            
\ref{17}{M. Weyrauch and H.J. Weber}{\PLB\ {\bf 171}, 13
(1986)};              
\refs{H.J. Weber and H.T. Williams}{\PLB\ {\bf 205}, 118
(1988)}.  
%
\ref{18}{L.Ya. Glozman and D.O. Riska}
{\NPA\ {\bf 603}, 326 (1996).} 
\ref{19}{Fl. Stancu}{Group Theory in Subnuclear Physics, Oxford University
Press, 1996, (in press), chapter 4;}
\refs {S. Pepin and Fl. Stancu}{ Preprint ULG-PNT-96-1-J.}
 \bye